\documentclass[10pt, natbib]{article}

\usepackage{amsmath,amssymb,amsthm}
\usepackage[round]{natbib}
\newtheorem{theorem}{Theorem}
\newtheorem{proposition}{Proposition}
\newtheorem{remark}{Remark}

\newcommand{\Beta}{\mathrm{B}}

\newcommand{\Alpha}{\mathrm{A}}

\usepackage{mathptmx}      % use Times fonts if available on your TeX system
\usepackage{parskip}
\usepackage[dvips]{graphicx}
\usepackage[export]{adjustbox}
\usepackage{enumerate}
\usepackage{mathrsfs}
\usepackage{longtable}
\usepackage{subfigure}
\usepackage[usenames]{color}
\usepackage{float}
\usepackage{caption}
\usepackage[space]{grffile}
\usepackage[english]{babel}
\usepackage{amsfonts}
\usepackage{tabto}
\usepackage{tabularx}
\restylefloat{table}
\usepackage{listings}
\usepackage{lstautogobble}
\usepackage{hyperref}
\hypersetup{colorlinks,citecolor=blue, linkcolor = blue,urlcolor  = blue}
\usepackage{newtxtext,newtxmath}
\usepackage{tikz}
\usepackage[autostyle, english=british, french=guillemets]{csquotes}
\usepackage{adjustbox}

\begin{document}

\title{Portfolio liquidation under transient price impact - theoretical solution and implementation with 100 NASDAQ stocks\thanks{This research has been supported through the profile partnership program between Humboldt-Universit\"at zu Berlin and National University of Singapore.The authors gratefully acknowledge the financial support of the Singapore Ministry
of Education Academic Research Fund Tier 1 at National University of Singapore.}}
%\titlerunning{Portfolio Liquidation and Market Microstructure}

\author{Ying Chen \footnote{Department of Mathematics and Risk Management Institute, National University of Singapore,
	Lower Kent Ridge Road 10, 119076 Singapore, Singapore; email: matcheny@nus.edu.sg},   Ulrich Horst\footnote{Department of Mathematics, and School of Business and Economics, Humboldt-Universit\"at zu Berlin
	Unter den Linden 6, 10099 Berlin, Germany; email: horst@math.hu-berlin.de}~ and Hoang Hai Tran \footnote{Department of Statistics \& Applied Probability, National University of Singapore, 6 Science Drive 2, Singapore 117546, Singapore; email: e0045275@u.nus.edu}}

\maketitle

\begin{abstract}
	We derive an explicit solution for deterministic market impact parameters in the \citet{Graewe:Horst:2016} portfolio liquidation model. The model allows to combine various forms of market impact, namely instantaneous, permanent and temporary. We show that the solutions to the two benchmark models of \cite{almgren2001optimal} and of \cite{Obizhaeva2013} are obtained as special cases. We relate the different forms of market impact to the microstructure of limit order book markets and show how the impact parameters can be estimated from public market data. We investigate the numerical performance of the derived optimal trading strategy based on high frequency limit order books of 100 NASDAQ stocks that represent a range of market impact profiles. It shows the strategy achieves significant cost savings compared to the benchmark models of \cite{almgren2001optimal} and \cite{Obizhaeva2013} .
			
	\mbox{ } 
			
	{\bf AMS Subject Classification:}  62M10, 62P20
			  
	{\bf Keywords:} Liquidity Risk, Optimal Trading Strategy, Portfolio Liquidation, Hawkes process
\end{abstract}

\section{Introduction}\label{sec1}

In a perfectly elastic market, all market orders can be instantaneously executed without time delay and without any impact on market prices. Empirical evidence
indicates that large trades, however, are often settled at `worse' prices than small trades, due to adverse market impact. In real situations, traders face liquidity risk from market order imbalance, price uncertainty, and market microstructure embedded in the modern high frequency electronic trading system. Often, immediate execution of a large order is either impossible or extremely expensive because of high instantaneous execution costs. Alternatively, large orders may be sliced into smaller chunks that are submitted over the day. This usually generates less immediate yet of longer-lasting market impact by driving prices in an unfavourable direction for the trader.  

Market impact models have long been studied in the economics literature. The focus of the economics literature on market impact is typically on the role of information asymmetries, inventory effects and/or liquidity competition and how they affect asset prices. Early work includes \cite{glosten1985bid}, \cite{EasleyOHara} and \cite{kyle1985continuous}.
% \footnote{More recently, the impact of a particular form of information asymmetry, namely the use of hidden liquidity, on optimal trade execution and liquidity competition has received considerable attention in the financial economics literature; see \cite{BoulatovGeorge2013}, \cite{ButiRindi}, \cite{CH} and  \cite{Monais}, among others.}  
In the wake of the dramatic increase in automation trading, problems of optimal trade execution have also received considerable attention in the financial mathematics and quantitative finance literature; see the textbooks by \cite{Cartea-book} and \cite{gueant2012optimal} and references therein. This line of research focuses mostly on frictions that purely arise from trade execution without endogenizing trading motifs. This is in line with the view of \cite{Foucault2005}, who stress that a considerable amount of trading frictions in practice is not caused by information asymmetry but by the process of optimal trade execution. In practice, the decision to trade is typically separated from the process of trade execution as most market participants delegate this task to specialized brokerage firms. At this stage trade execution follows generic principles of transaction cost minimization irrespective of the investor's underlying trading motif, and optimal trading strategies are computed for exogenous market impact functions (as estimated from transaction data or order book dynamics), rather endogenously derived ones (e.g. through equilibrium approaches).  

Market impact is typically classified as {\textsl instantaneous, permanent} or {\textsl temporary (persistent)}. With instantaneous impact a trader receives a price only instantly deviating from the current price available; permanent impact increases all future transaction prices in the same way; temporal impact vanishes after the execution of the order as time passes. For linear instantaneous and linear permanent price impact, \cite{bertsimas1998optimal} derived dynamic optimal trading strategies for a risk-neutral investor based on the minimization of the expected cost of execution. \cite{almgren2001optimal} extended this model to risk-averse investors and gave a closed-form solution for the optimal execution strategy in a mean-variance framework. \cite{Huberman:Stanzl:2005} showed that linear functions are the only choice of the permanent price impact for which the model is free from arbitrage. 

A second line of research initiated by \cite{Obizhaeva2013} assumes that there is instantaneous and temporary impact with the impact of past trades on current prices decaying over time. The absence of arbitrage in the sense of Huberman and Stanzel in linear-impact continuous-time impact models with transient price impact was established in \cite{GatheralSchiedSlynko}. Optimal trading strategies in the Obizhaeva-Wang modelling framework usually comprise large block trades at the beginning and at the end of the trading period while in models with instantaneous and permanent impact usually only continuous strategies are allowed; see \cite{HorstNaujokat} for further details. In many instances, the block trades exceed the liquidity available for trading by several orders of magnitude for realistic choices of impact parameters. 

\cite{Graewe:Horst:2016} proposed a unified probabilistic framework for analyzing optimal liquidation models with both instantaneous, permanent and temporary market impact using only continuous strategies. This avoids the problems resulting from block trading while maintaining the realistic feature of decaying price impact. They established the existence of a unique optimal trading strategy and showed that the optimal strategy can be  characterized in terms of the solution to a complex system of backward-stochastic differential equations with singular terminal condition. When all impact parameters are deterministic, the stochastic equation reduces to a system of ordinary differential equations (ODE). We provide an explicit solution for deterministic impact parameters using a calculus of variations approach, and show (i) how the different forms of market impact relate to the microstructure of limit order book markets and (ii) how the impact parameters can be estimated from public market data. 

The limit order book contains the key features of market microstructure and can be used to describe important factors that determine price impact. There is a significant economic and econometric literature on order books including \cite{Biais}, \cite{EasleyOHara}, and \cite{glosten1985bid} that emphazises on the realistic modelling of the working of the order book, and on its interaction with traders' order submission strategies. More recently, a series of high-frequency limits for structural order book models has been established in financial mathematics literature; see~\cite{AbergelJedidi2011}, \cite{Cont1} and \cite{HP} and references therein. At high frequency, the order book provides comprehensive statistical characteristics of the underlying variables such as price resilience and order arrivals. We interpret the different forms of market impacts in terms of order flow dynamics. Instantaneous impact depends on market depth through limit order arrivals and cancellations; permanent and temporary impact are triggered by the shift in the mid-price process, due to self-exciting anticipation of future order flows. 

%Several theoretical and empirical studies have documented self-exciting and clustering behaviour in financial markets. 
Self-exciting dynamics can be naturally modelled by the Hawkes processes. Originally introduced in \cite{Hawkes1971}  to model the occurrence seismic events, Hawkes processes have recently received considerable attention in the financial mathematics and economics literature as a powerful tool to model financial time series dynamics. Their application range from trade arrivals in high-frequency markets \citep{SahaliaCacho-DiazLaeven2015,AndersenFusariTodorov2015,Bacry2012,HLR2015}, to 
%dark pool trading \citep{Gao2018,LehalleBook2018} and 
volatility modelling \citep{Bates2019,ElEuchRosenbaum2019a,HorstXu2019b}, and from limit order book modelling \citep{HorstXu2019a} to market impact and microstructure \citep{Alfonsi2016,bacry2015,cartea2011buy}.

We also brige the gap between theoretical models and implementaiton. Estimation of market impact factors often relies on the availability of proprietary dataset. For instance, \cite{almgren2005direct} uses a proprietary dataset to regress the market impact factors in the \cite{almgren2001optimal} model. \cite{fraenkle2011market} requires a proprietary dataset from real trading to estimate market impact of a VWAP trading strategy. It is more desirable to be able to extract the relevant market impact factors from public market data, making an easy-to-replicate estimation. \cite{Jondeau2008} used tick-by-tick data of one French stock, Orange, traded in the Paris Stock Exchange to estimate the parameters in the price process. \cite{Mon2005} used tick-by-tick data of the Helsinki Stock Exchange to estimate the parameters of the model. These works, however, do not estimate the full spectrum of market impact, namely the resilience of market impact. We propose a statistical procedure to estimate the unknown market factors from the public order book data, making the implementation of the order splitting strategy practically possible for every stock.

Our contributions are summarized as follows. First, we derive a closed form solution or the liquidation problem in a unified framework that allows for various forms of market impact. It should be noted that our derivations are different from the work of \cite{Obizhaeva2013}. Their optimal solution comprises mixed strategies from both discrete and continuous trades, while we only consider continuous strategies, which allows comparison of optimal strategies in the same class of continuous trading strategies. Second, as far as we are aware, our work is the first one deriving and connecting all the three impacts in a unified framework and to relate the various forms of market impact, especially the temporary impact to order flow dynamics. This provides a method to estimate the parameters required to implement the order-splitting strategies for real data. As illustration, we calibrate the market impacts of $100$ NASDAQ stocks that represent a wide range of profiles. Numerical results show that among the $100$ stocks
\begin{itemize}
	\item The instantaneous market impact factors have an average value of $0.022$ bps per share, with relatively large variation, presumably due to liqudity availability for each stock;
	\item The permanent market impact factors has an average value of $0.008$ bps per share, with relatively tighter range than the instantaneous market impact factors;
	\item The permanent market impact is expected to lessen by half over on average of $0.608$ day.
\end{itemize}

The proposed model also enables an insightful numerical comparison on execution cost among different optimal strategies based on real data. Given the estimators of the three types of market impact and the optimal trading strategy, we study the performance of different optimal order-splitting strategy by calculating the execution costs from different order-splitting strategies. Our numerical results show that the derived strategy achieves significant cost savings compared to the alternative optimal strategies.

The paper is organized as follows: Section \ref{secMethod} describes the dynamics of market impact factors and how to calibrate them based on market microstructure dynamics. Section \ref{secRealDataAnalysis} reports the estimation of the market impacts with real market data. Section \ref{secStrategy} investigates numerical performance of different optimal strategies in terms of execution costs. Section \ref{secConclusion} concludes.

%%%%%%%%%%%%%%%%%%%%%%%%%%%
%%%%%%%%%%%%%%%%%%%%%%%%%%%
%%%%%%%%%%%%%%%%%%%%%%%%%%%

\section{Optimal portfolio liquidation and market microstructure}\label{secMethod}

In this section, we present the portfolio liquidation model introduced in \cite{Graewe:Horst:2016} and derive an explicit solution for deterministic market impact parameters. The models combines instantaneous, permanent and temporary impact into a unified framework. We show how the solutions to the two benchmark models of \cite{almgren2001optimal} and of \cite{Obizhaeva2013}  are obtained as special cases. Subsequently we show how the different forms of market impact relate to the microstructure of limit order book markets. 

\subsection{The liquidation model}

We consider a market model where the dynamics of benchmark price process is described by some continuous-time stochastic process $S=(S_t)_{t \in [0,T]}$ defined on a filtered probability space $(\Omega, {\cal F}, ({\cal F}_t)_{t \in [0,T]}, \mathbb{P})$. As it is customary in the liquidation literature we assume that $S$ is a continuous martingale.

We consider a risk neutral investor who needs to liquidate a large position of $x > 0$ shares of some stock over the time period $[0,T]$. Trading takes place in continuous time. Following the majority of the portfolio liquidation literature we assume that the trader liquidates the portfolio using only absolutely continuous trading strategies. A trading strategy is an adapted stochastic process $\xi: \Omega \times [0,T] \to \mathbb{R}$ where $\xi_t$ denotes the rate at which the trader sells the stock at time time $t \in [0,T]$. Associated with the strategy $\xi$ is the portfolio process
	\[
		X_t = x - \int_0^t \xi_s ds, \quad t \in [0,T]
	\]
that describes the dynamics of the trader's stock holding. A trading strategy is called admissible if it is square integrable with respect to the measure $dt \otimes \mathbb{P}$ and satisfies the liquidation constraint $$X_T=0.$$

We allow for instantaneous, permanent, and temporary market impact. Instantaneous impact adds an immediate liquidity cost to each trade but does not affect the cost of future trades. The instantaneous impact parameter is denoted $\eta >0$. Permanent market impact adds a drift to the fundamental price process and affects the transaction cost of all future trades in the same way.  The permanent market impact parameter is denoted $\lambda >0$. Temporary impact also adds a drift to the fundamental price process but affects the cost of future trades at a decreasing rate. Denoting the temporary impact factor by $\gamma > 0$ and the recovery rate by and $\rho > 0$, the dynamics of the temporary price impact associated with a trading strategy $\xi$ is described by the process
\[
	Y_t = \int_0^t \left\{ \gamma \xi_s - \rho  Y_s \right\} ds, \quad t \in [0,T].
\]
It is important that $Y_0 = 0$; otherwise, the model might allow a form of arbitrage as illustrated in \cite{Graewe:Horst:2016}. The different forms of market impact result in the following execution price process for the large trader:
\[
	P_t = S_t -  \eta \xi_t - \lambda \int_0^t \xi_s ds  - Y_t, \quad t \in [0,T].
\]
The trading costs are given by the expected liquidation shortfall, that is by the expected difference between the book value and trading gains as
\[
	\mathbb{E}\left( x S_0 - \int_0^T P_t \xi_t dt  \right).
\]
The expected shortfall can be expressed in terms of the instantaneous, the permanent and the persistent market impact processes defined by
\[
	E_t := \eta \int_0^t \xi^2_s ds, \quad G_t := \lambda \int_0^t \xi_s ds:=\lambda(x_0-X_t),  \quad F_t := \int_0^T \xi_s Y_s ds
\]
respectively, where $t \in [0,T]$. We measure impact cost in basis points. Since the total permanent market impact is independent of the trading strategy it does not affect the optimal trading strategy and can hence be dropped from the optimization problem. Using the martingale property of the benchmark price process, the total expected execution cost associated with an admissible trading strategy is then computed as 
\[
	C_T := \mathbb{E} \left( E_T + F_T \right).
\]
The goal of the trader is to minimize the expected execution cost over all admissible trading strategies. It follows from \cite{Graewe:Horst:2016} that a unique optimal trading strategy $\xi^*$ exists and that $\xi^*$ is deterministic because we assume that the impact parameters are deterministic. The following theorem, which is proved in the appendix, provides an explicit representation of the optimal trading strategy.

\begin{theorem}\label{main1}
	The optimal solution to the portfolio liquidation problem with both instantaneous, permanent and temporary market impact is given by the deterministic trading strategy
	\begin{equation} \label{opt}
		\xi^*_t := \frac{k x_0 \left(\left(k^2 \eta-\gamma \rho\right) \left(\rho \cosh \left(\frac{k T}{2}\right)+k \sinh
			\left(\frac{k T}{2}\right)\right)+\gamma \rho^2 \cosh \left(k
			\left(t-\frac{T}{2}\right)\right)\right)}{k \rho T \cosh \left(\frac{k T}{2}\right)
			\left(k^2 \eta-\gamma \rho\right)+\sinh \left(\frac{k T}{2}\right) \left(\gamma \rho \left(2 \rho-k^2
			T\right)+k^4 \eta T\right)}
	\end{equation}
	and the resulting optimal portfolio process is given by
	\begin{eqnarray*}
		X^*_t
		%&=& x_0-\frac{x_0 \left(k \rho t \cosh \left(\frac{k T}{2}\right) \left(k^2 \eta-\gamma \rho\right)+\sinh
		%\left(\frac{k T}{2}\right) \left(\gamma \rho \left(\rho-k^2 t\right)+k^4 \eta t\right)+\gamma \rho^2 \sinh
		%\left(k \left(t-\frac{T}{2}\right)\right)\right)}{k \rho T \cosh \left(\frac{k
		%T}{2}\right) \left(k^2 \eta-\gamma \rho\right)+\sinh \left(\frac{k T}{2}\right) \left(\gamma \rho \left(2
		%\rho-k^2 T\right)+k^4 \eta T\right)} \\
		&=& x_0-x_0\left(\frac{a+bt+c\sinh(k(t-\frac{T}{2}))}{2 a+bT} \right),
	\end{eqnarray*}
	where
	\begin{eqnarray*}
		k &:=& \sqrt{\rho(\rho+\frac{\gamma}{\eta})}\\
		a &:=& \sinh\left(\frac{k T}{2}\right)\frac{\gamma}{\eta}\\
		b &:=& k\rho \cosh \left(\frac{k T}{2}\right) + k^2\sinh \left(\frac{k T}{2}\right)\\
		c &:=& \frac{\gamma}{\eta}.
	\end{eqnarray*}
\end{theorem}

One notable result is that the impact ratio $\frac{\gamma}{\eta}$ and the resilience parameter $\rho$ determine the shape of the strategy, rather than the absolute values of the market impact factors $\gamma$ and $\eta$ individually. A second notable result is that the optimal trading strategy is always positive. That is, when liquidating a single stock, it is not optimal to buy the stock at any point in time within this modelling framework. This is \textsl{not} always the case when impact parameters are stochastic or multi-asset portfolios are liquidated. 

The benchmark model of \cite{almgren2001optimal} with only instantaneous impact corresponds to the special case $\gamma = 0$. In this case, the optimal liquidation strategy consists of a TWAP strategy.  When considering both instantaneous and temporary market impacts, the optimal strategy becomes a weighted average of a TWAP strategy and a V-shaped strategy. The solution to the benchmark model of \cite{Obizhaeva2013} is obtained as the limiting case when the instantaneous impact tends to zero. This shows that our model interpolates between the two probably most investigated liquidation models. More precisely, we have the following result. The proof is straightforward and hence omitted.

\begin{proposition}\label{main2}
	\begin{itemize}
		\item[i)] If $\gamma = 0$, then $X^*_t = x - x \frac{t}{T}$, for any $t \in [0,T]$.
		\item[ii)] If $\frac{\gamma}{\eta} \to \infty$, then for any $t \in [0,T]$,
		      \[
		      	X^*_t \to x_0 - \frac{x_0}{\rho T +2}(H_0(t) + \rho t + H_T(t))
		      \]
		      where $H_a(x) = \left\{ \begin{array}{ll} 1 & \mbox{ if } x \geq a \\ 0 & \mbox{ else } \end{array} \right.$ is the Heaviside function in $a \in \mathbb R$.
	\end{itemize}
\end{proposition}

%%%%%%%%%%%%%%%%%%%%%%%%%%%%
%%%%%%%%%%%%%%%%%%%%%%%%%%%%
%%%%%%%%%%%%%%%%%%%%%%%%%%%%

\subsection{Market microstructure and market impact}\label{sec-micro}

We proceed to provide an interpretation of our market impact parameters in terms of order flow dynamics and show how the parameters can be estimated from public market data. In particular, we show how the permanent and temporary are triggered by a shift in the mid-price process, due to anticipation of future order flows.

To this end, we consider a queuing-theoretic order book model similar to \cite{Cont1} and \cite{HP} with constant spread $p_0$ from the mid price; most liquid stocks are traded at a fixed spread, usually with one tick. The tick size is denoted $\Delta$ and $p_i := p_0 + i \Delta$ $(i=1,2, ...)$ denotes the price level $i$ ticks into the bid side of the book. We take the mid-price process as our benchmark price process. In the absence of market impact, mid-price changes occur according to an exogenous point process; they are monitored at a rate $\mu$. Conditioned on a price change, an up movement occurs with probability $p^{up}$ and a down movement occurs with probability $p^{down}$. The mid-price process is a martingale if $p^{down} = \frac{1}{2}$. Since order imbalances have been identified as important determinants of price movements \citep{Biais,CH,cont2010price} we will condition the probabilities of up and down movements on market order imbalances; this will generate the permanent impact.  

We also assume that buy and sell side market orders occur according to independent Hawkes processes with intensity processes $\nu^b$ and $\nu^s$ that will depend on the large trader's liquidation strategy. The additional child resulting from the dependence of the Hawkes dynamics on the large trader's strategy will generate the temporary price impact.

Finally, we assume that limit orders are placed, respectively cancelled according to independent Poisson processes at rates $\mu^+_i$, respectively $\mu^-_i$ at the price level $p_i$. The corresponding order sizes are modelled by independent and identically distributed random variables $V_i$ with means $\bar V_i$, $(i=0,1, ...)$. In between mid-price changes the volume dynamics across different price levels follow independent Markov processes whose distributions depend on the distance to the best price. The considerable empirical evidence that limit order arrivals and cancellations occur at much higher rates than mid-price changes suggests to view the order book as set of independent infinite-server queues, and to model the number of orders standing at the $i$-th price level by the stationary distribution of the respective $M/M/\infty$ queuing model. The stationary distribution at the $i$-th level is the Poisson distributions with rate $\frac{\mu_i^+}{\mu_i^-}$ and the expected number of shares at that level is $\bar q_i := \bar V_i \frac{\mu_i^+}{\mu_i^-}$. 

Under the stationarity assumption the distribution of the liquidity relative to the best price is independent of the mid-price. This corresponds to a situation where market makers place their orders at fixed distances from the mid-price. \cite{avellaneda2008high} provides a theoretical foundation for such strategies, where a market maker's bid-ask quote is a spread function maximizing her utility upon obtaining one additional share given her current inventory, market order arrival rates and risk aversion factor. Assuming no inventory, a market maker's bid-ask quote remains a fixed constant around the mid-price.

\subsubsection{Instantaneous impact}

We interpret that the instantaneous market impact is a direct result of market makers' demand for carrying additional inventory. From both empirical and theoretical
standpoints, rational market makers demand a liquidity premium for each limit orders posted to compensate for being adversely selected and/or hedging cost against price volatility. For example, in \cite{madhavan1997security}, the bid-ask spread has a linear component representing market makers' compensation. This is also consistent with the interpretation of instantaneous market impact given in \cite{almgren2001optimal}.

In order to quantify the instantaneous execution cost we consider the LOB in the high frequency limit and denote by $\bar Q_i := \sum_{j=0}^i \bar{q}_j$ $(i=0,1, ...)$ the expected cumulative volume distribution function. This function can be viewed as an average inverse liquidity offer curve. Assuming small tick sizes so that prices are almost continuous quantities it is convenient to work with the continuous linear inverse offer curve
\[
	p(x) = p_0 + \eta x, \quad x \geq 0
\]
that is given by the best linear approximation to the expected (discrete) volume distribution function. The expected instantaneous execution price for executing $x$ shares is then given by $p=p_0+ \eta x,$ and the expected cost of instantaneous execution is $\eta x_0^2.$ We notice that our high-frequency regime is consistent with the idea that the order book recovers rapidly from larger trader submissions. 

%%%%%%%%%%%%%%%%%%%%%
%%%%%%%%%%%%%%%%%%%%%
%%%%%%%%%%%%%%%%%%%%%

\subsubsection{Permanent impact}

We consider {permanent market impact} as a permanent drift being added to the fundamental price process after each trade, resulting in a linear cost throughout the execution. There are two streams of explanation for this phenomenon. The first one, pioneered by \cite{kyle1985continuous} assumes that the linear market impact is due to informed traders revealing their private information on future price movements through his trading activities. A second proposition, followed in, e.g.~\cite{Jaisson2015} assumes that the market impact is due to imbalance in supply and demand and therefore creates a deviation from an otherwise martingale price process. A subtle difference between the two interpretations is that the act of trading does not affect the price process in the first one, but merely revealing information about it over time, while in the second one, the price impact is the direct result of trading. We follow the second interpretation.

Let us assume that market buy and sell orders arrive according to (conditionally) independent point processes with respective intensity processes $\nu^b$ and $\nu^s$. The precise dynamics will be specified in the next subsection. We define
\[
	\delta_t := \int_0^t \left( \nu^s_s - \nu^b_s \right) ds
\]
as the expected order imbalance at time $t \in [0,T]$. Accounting for the dependence of price dynamics on order imbalances, we assume that the probability of a down  movement at time $t \in [0,T]$, conditioned on a mid-price change taking place at that time is given by
\begin{equation} \label{def-f}
	p^{down}_t = f\left( \delta_t \right)
\end{equation}
for some differentiable function $f: \mathbb{R} \to [0,1]$. The mid-price process is a martingale if $p^{down}_t \equiv \frac{1}{2}$, i.e. if
\[
	\nu^s_t - \nu^b_t \equiv f^{-1}\left( \frac{1}{2} \right) =: \bar \delta.
\]
The quantity $\bar \delta$ can be viewed as the equilibrium imbalance between market sell and buy order arrivals.

We assume that the unaffected mid-price process follows a martingale and that a single sell order placed by the large trader changes the probability of the next price movement to be a down movement to 
\[
	p^{down} = f\left( \bar \delta + 1 \right) \approx \frac{1}{2} + f'\left( \bar \delta \right).
\]

Let $\bar{L}$ be the average size of a market order, and let $\bar Z$ be the average magnitude of a price change resulting from an average-sized market order submission. A trading strategy that submits average sized market sell orders at a constant rate $\xi \leq \mu$ increases the down-movement probability of $100\frac{\xi}{\mu}$\% of all price changes\footnote{We later choose $\mu \equiv 1$.}  and hence adds a constant drift $-2 \frac{\xi}{\mu} f'(\delta) \bar Z$
%\[
%	\left( 2 \frac{\xi}{\mu} f'(\delta) \bar Z t \right)_{0 \leq t \leq T}
%\]
to the unaffected benchmark process. We thus define the impact per average size market order, respectively share as
\[
	\Lambda := 2 \frac{1}{\mu} f'(\delta) \bar Z \quad \mbox{and} \quad \lambda := 2 \frac{1}{\mu} f'(\delta) \frac{\bar Z}{\bar L}.
\]
For the special case of TWAP schedule submitting average size market orders at a rate $\xi = \frac{x}{T \bar L}$, where $x$ denotes the initial portfolio, adds the constant drift $-\frac{\lambda}{\mu} \frac{x}{T}$ 
%\[
%	\left( \frac{\lambda}{\mu} \frac{x}{T} t \right)_{0 \leq t \leq T}
%\]
to the unaffected benchmark price process.

%%%%%%%%%%%%%%%%%%%%%%%
%%%%%%%%%%%%%%%%%%%%%%%
%%%%%%%%%%%%%%%%%%%%%%%

\subsubsection{Temporary impact}

We interpret temporary market impact as the expectation of future permanent market impact, due to the persistence of trade flows. Empirically, it has been observed that the sign process of the market orders is highly persistent. For example, \cite{Bouchaud2004} has shown that the sign process of market orders in France-Telecom stock market reveals very slowly decaying correlations as a function of trade time. There are at least two different explanations for this clustering effects in order flows. The first is a herding effect, where small investors follow institutional orders to trade. This is normally attributed to the asymmetric information between informed and noise traders, in which the latter imitates the former instead of forming their own actions. Supporting empirical evidence for herding effects is given in, e.g. \cite{ZHOU2009388}, where the herding behaviour is found stronger in small cap stocks against large cap and larger in Hong Kong Composite Index than in Mainland Composite Index using Hong Kong Stock Exchange (HKEX) data.

A second possible explanation that has attracted increasing attention is splitting effects, where each parent-order is split into multiple child-orders before routed into the market.
Using a proprietary dataset of London Stock Exchange (LSE), in which each trade is tagged with a membership code indicating the originating broker, \cite{BOUCHAUD200957} 
found that the autocorrelation among same broker's trades is at least an order of magnitude higher than the rest. It has been concluded that the splitting effect being able to reflect the self-exciting and clustering features, instead of the herding effect, is the dominating reason of the long memory in order flow.

Regardless of the cause, it has been widely recognized that there a long-term memory effect in market order flow, and the Hawkes process has been suggested to model this effect; e.g.~\cite{bacry2015} and \cite{Hawkes2018} among others. A Hawkes process with exponential kernel $A e^{-Bt}$ is a point process $N:\Omega \times [0,T] \to \mathbb{N}$ with an intensity process of the form
%\begin{equation} \label{intensity-Hawkes}
%	I_t = \nu + \int_0^t A e^{-B (t-s)}N(ds), \quad t \in [0,T]
%\end{equation}
\begin{equation} \label{Hawkes-intensity}
	\nu_t = \nu + \int_ 0^t A e^{-B(t-s)} dN(ds).
\end{equation}
where $\nu$ is the base intensity, and the integral term captures the self exciting effect of event arrivals. The ratio $\frac{A}{B}$ is called the branching ratio. The stationary condition for the Hawkes process is $\frac{A}{B}<1$.

In the preceding subsection we only assumed that market buy and sell orders arrive at an equilibrium rate. In this section we specify their arrival dynamics in a way that allows us to link temporary market impact to the clustering of order flow.\footnote{Our interpretation is different from \cite{Obizhaeva2013}, in which the temporary market impact is due to the temporary pressure on the order book, which recovers slowly overtime.}
Specifically, we assume that the unaffected buy and sell side market order flows follow independent Hawkes processes with the intensity processes (\ref{Hawkes-intensity}).
%\begin{equation} \label{Hawkes-intensity}
%	\mu_t = \mu + \sum_{i=1}^{N_t} \int_ 0^t A e^{-B(t-s)} dN(ds).
%\end{equation}
Assuming the trader submits additional market sell orders at rate $\xi$, it increases the sell side base intensity to $\nu + \xi$ and hence changes the arrival dynamics of sell orders to a Hawkes process $N$ with intensity
\[
	\bar \nu_t = \nu + \xi + \int_ 0^t A e^{-B(t-s)} dN(ds).
\]

The temporary market impact factor can be interpreted as the permanent market impact caused by the additional child-orders originating from the large trader's activity, due to herding effects. To simplify the analysis we assume that only the respective first generation child-orders have market impact. This means that only direct descendant orders caused by the large trader activities are due to herding effects, while the rest is due to splitting effects that do not cause additional market impact. From the cluster representation of Hawkes processes we conclude that the expected number of first generation child-orders originating from an order placed at time $t_i$ is given by

$$\int_{t_i}^T  A e^{-B (T-t)} dt= \frac{A}{B} (1-e^{-B \left(T-t_i\right)}).$$

Similar to the previous subsection we benchmark our temporary market impact factor on a TWAP strategy that executes $\bar{L}$ number on each transaction and hence consider the constant rate $\frac{x}{T \bar L}$ where $x$ is the initial portfolio size. In this case, the total expected additional impact is
\[
	\frac{\lambda}{\mu} \frac{x}{T} \frac{A}{B} \int_0^T \left( 1-e^{-B \left(T-t\right)} \right) dt = \frac{\lambda}{\mu} x \frac{A}{B} \left(1 + \frac{1}{T B} (1-e^{-BT}) \right).
\]
Equating the above quantity with the terminal value $Y_T$ of the temporary market impact process corresponding to a continuous time TWAP schedule and dropping lower order terms, we obtain that
\[
	\frac{x \gamma}{T} \int_0^T  e^{-\rho (T - s)} ds = \frac{x \gamma}{T \rho}\left( 1 - e^{-\rho T} \right) = \frac{\lambda}{\mu} x \frac{A}{B}.
\]
Assuming that $\frac{\lambda}{\mu}=\gamma$, that is, the initial shock of the temporary market impact factor is due to the permanent market impact, and using a Taylor approximation of the exponential function up to the second order we obtain that
\[
	\rho \approx \frac{1-A/B}{T/2}.
\]
This shows how resilience factor links to the self-exciting dynamics dynamics of market order flow. 

\begin{remark}
	We emphasise that the resilience parameter was derived for a TWAP schedule. Different strategies will naturally lead to different resiliences. We shall nonetheless use this easy-to-compute parameter as an input parameter for the general liquidation model and our empirical analysis.
\end{remark}

%\begin{remark}
% We notice that we computed the resilience factor for the benchmark of a TWAP strategy; other strategies will result in other resilience factors.  {\bf We need to justify that we use THIS particular value for our optimal strategy.} 
%\end{remark}

%%%%%%%%%%%%%%%%%%%%%%%%%%%%%%%
%%%%%%%%%%%%%%%%%%%%%%%%%%%%%%%
%%%%%%%%%%%%%%%%%%%%%%%%%%%%%%%
%%%%%%%%%%%%%%%%%%%%%%%%%%%%%%%

\section{Data, statistical approaches and impact calibration}\label{secRealDataAnalysis}

In this section, we will present the public order book dataset over 100 NASDAQ stocks. We will detail the statistical approaches for estimating the market impact factors based on real data and illustrate the estimation using the Intel Corporation's stock (NASDAQ: INTC) as example. Lastly we will provide statistical summary of the estimated parameters for all the stocks.

\subsection{Data description}
Our dataset is from National University of Singapore, which offers reconstructed limit order book data derived from NASDAQ's Historical TotalView-ITCH files. For every order data set, it is comprised of all the streaming messages from exchanges. Each message in the dataset contains the event timestamp, message type, order id, order size, order price and trade direction. There are six types of events, namely submission of a new limit order, modification of a limit order, cancellation of a limit order, execution of a visible limit order, execution of a hidden limit order and trading halt indicator. Since our order book model only considers submission, cancellation and execution of limit orders, we strip the rest of the events from the dataset. We also classify the messsages further to identify the order book levels and perform the analysis of the liquidity of each level. {We conduct estimation at second frequency, that is we assume $\mu=1$ in the model introduced in Section \ref{sec-micro}. At this particular frequency, empirical features of real data are consistent to the model assumptions as well as the statistical approaches to be used in calibration such as the logistic regression for permanent impact factor. Moreover, this choice avoids the impact of microstructure noise compared to possibly ultra-high frequency such as millisecond in the raw data.} Further to avoid the spike of volatility nearing the opening and closing auctions, we use the data from 10:00 a.m to 3:30 p.m Eastern Time, which is the period of 30 minutes after the opening auction and 30 minutes before the closing auction of NASDAQ exchange.

To ensure a comprehensive study, we pick 100 stocks across NASDAQ listed companies that represent a wide range of market sectors, including Technology, Public Ultilities, Finance and Health Care; and market capitalization, from 6 billion USD (LEG) to more than 800 billion USD (MSFT), and also in terms of number of events, from nearly 29,000 events a day for J M Smucker Co (SJM) to more than 1 billion events a day for Intel Corporation (INTC). The details can be found in the Appendix.

As an illustration, we report the statistics for INTC and proceed to estimate the parameters based on the order books of INTC. We consider the 4th of April, 2018, a normal Wednesday without any major events, representing normal trading activities. The summary statistics for the first 10 order book levels for INTC is reported in Table \ref{table:ordStat}. There is a decreasing count of submission and cancellation events over levels, implying the activity reduction of liqudity providers as the price moves away from mid-price. Similarly, the decreasing count of the execution events show market orders tend to hit the near mid-price levels more often, but there is non-zero activity counts on higher levels, showing that there are still orders that hit multiple levels at once. This is consistent with our assumptions in the derivation of instantaneous market impact.

\begin{center}
	\begin{table}[!htb]
		\centering
		\begin{tabular}{c|r|r|r|r|r|r|r|r|r|r}
			\hline
			\hline
			Lvl & 1       & 2      & 3      & 4      & 5      & 6      & 7     & 8     & 9     & 10    \\
			\hline
			Sub & 344,508 & 78,117 & 30,363 & 21,705 & 15,964 & 11,847 & 7,584 & 8,034 & 8,981 & 6,335 \\
			\hline
			Cxl & 323,899 & 75,131 & 29,696 & 21,358 & 15,766 & 11,658 & 7,468 & 7,973 & 8,856 & 6,259 \\
			\hline
			Exc & 20,575  & 2,964  & 655    & 339    & 190    & 180    & 107   & 59    & 115   & 74    \\
			\hline
			\hline
		\end{tabular}
		\caption{Event counts per level of the order book for INTC, which shows the decreasing submission and cancellation rate, resulting a similar expected number of shares similar to our assumptions.}
		\label{table:ordStat}
	\end{table}
\end{center}

%%%%%%%%%%%%%%%%%%%%%%%
%%%%%%%%%%%%%%%%%%%%%%%
%%%%%%%%%%%%%%%%%%%%%%%

\subsection{Calibration of order-splitting strategies for a single stock INTC}

We proceed to estimate the parameters in the order-splitting strategies and calculate the resulting trading costs.  We also show the empirical features of the market data and discuss their consistency with our model assumptions.

\subsubsection{Instantaneous impact factor}\label{exmpl_inst}
Instantaneous market impact is an instant cost of placing a trade that leads to immediate recovery. We estimate instantaneous trading costs assuming a constant spread $p_0$ from the mid price. This is a well justified assumption for liquidly traded stocks. In real order books the number of shares standing at a particular price level may be zero. Let $P_i$ be the offer price at the $i$-th price level with positive liquidity, and $O_i=P_i-p_0$ be the corresponding price offset from the mid-price. We propose to regress $\eta$ as the slope against the empirical average values on each price level. Specifically, let $\tilde{\mu}^{\pm}_i$ and $\tilde{V}_i$ be the empirically observed order arrival/cancellation rates and the empirically observed average order sizes at the $i$-th level from the mid-price. Here we adopt $M/M/\infty$ queuing model. Moreover, given the fact that the market order arrival rate is much smaller than the limit order arrival and cancellation rates, as reflected in Table \ref{table:ordStat}, we approximate liquidity provided at each level by the order arrival/cancellation rates only. Let $\tilde{Q}_i:=\sum_{j=0}^i (\frac{\tilde{V_j} \tilde \mu^{+}_j}{\tilde \mu^{-}_j})$ be the empirically observed liquidity distribution function. Then $\hat{\eta}$, the point estimator of $\eta$, can be calculated as the least square estimator of the regression equation $$O_i = p_0 + \eta \tilde Q_i + \epsilon_i$$ where the residuals $\epsilon_i$ are random variables drawn independently from some distribution with mean zero. Empirical arrival rates and average order sizes along with the relevant derived metrics for a buying strategy of INTC are reported in Table \ref{table:LimitPoissonParameters}. We estimate the instantaneous market impact factor per share in ticks as $0.0011$. If converted to basis points, based on the opening price of $48.625$ we find that $$\eta=0.00226\hbox{ basis points}.$$

\begin{table}[]
	\centering
	\begin{tabular}{r|r|r|r|r|r|r}
		\hline
		\hline
		\multicolumn{1}{c|}{Lvl} & \multicolumn{1}{c|}{$\mu^{+}$} & \multicolumn{1}{c|}{$\mu^{-}$} & \multicolumn{1}{c|}{$\bar{V}$ (shares)} & \multicolumn{1}{c|}{$q_i$ (shares)} & \multicolumn{1}{c|}{$Q_i$ (shares)} & \multicolumn{1}{c}{$O_i$ (ticks)} \\ \hline
		1                        & 8.151                          & 1.008                          & 108.665                                 & 878.623                             & 878.623                             & 1                                 \\ \hline
		2                        & 2.146                          & 0.242                          & 133.250                                 & 1178.778                            & 2057.401                            & 2                                 \\ \hline
		3                        & 0.965                          & 0.205                          & 417.150                                 & 1960.409                            & 4017.810                            & 3                                 \\ \hline
		4                        & 0.704                          & 0.247                          & 776.340                                 & 2214.474                            & 6232.284                            & 4                                 \\ \hline
		5                        & 0.423                          & 0.101                          & 113.832                                 & 479.835                             & 6712.119                            & 5                                 \\ \hline
		6                        & 0.291                          & 0.094                          & 94.858                                  & 293.322                             & 7005.441                            & 6                                 \\ \hline
		7                        & 0.201                          & 0.126                          & 118.644                                 & 189.527                             & 7194.968                            & 7                                 \\ \hline
		8                        & 0.219                          & 0.101                          & 102.234                                 & 220.446                             & 7415.414                            & 8                                 \\ \hline
		9                        & 0.213                          & 0.055                          & 109.862                                 & 421.166                             & 7836.580                            & 9                                 \\ \hline
		10                       & 0.137                          & 0.111                          & 106.525                                 & 132.373                             & 7968.954                            & 10                                \\ 
		\hline
		\hline
	\end{tabular}
	\caption{Data to calculate instantaneous market impact factor based on INTC's data. $q_i=\bar{V_i}  \frac{\mu^{+}_i}{\mu^{-}_i}$ (shares) and $Q_i=\sum_{i=0}^N q_i$ (shares).}
	\label{table:LimitPoissonParameters}
\end{table}

%%%%%%%%%%%%%%%%%%%%%%%
%%%%%%%%%%%%%%%%%%%%%%%
%%%%%%%%%%%%%%%%%%%%%%%

\subsubsection{Permanent impact factor}\label{exmpl_perm}

\begin{figure}[!htbp]
	\begin{center}
		\includegraphics[width=\textwidth]{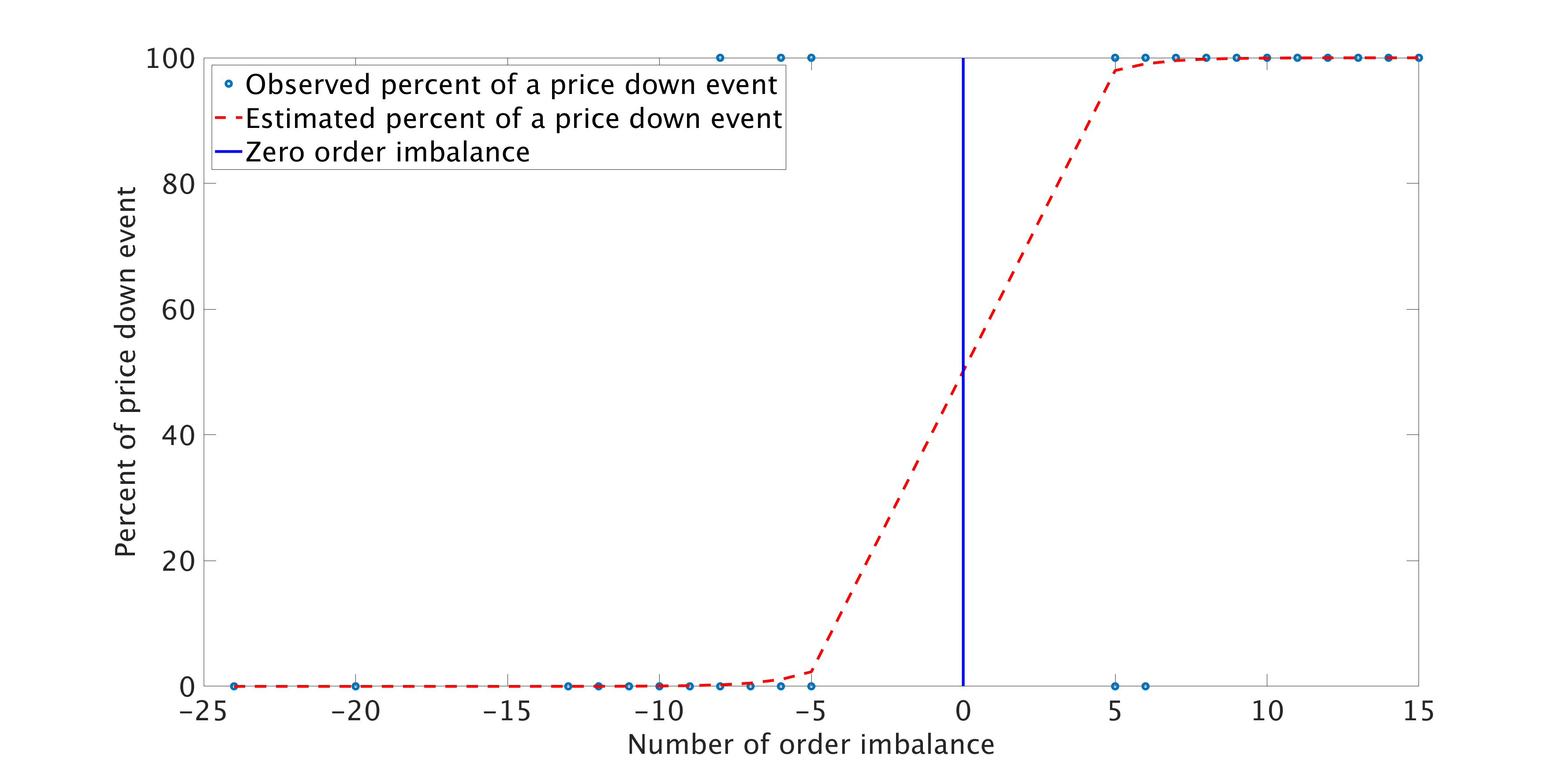}
	\end{center}
	\caption{Logistic regression model fit for change in mid-price as a result of market order imbalance. }
	\label{fig:logisticFit}
\end{figure}

Permanent market impact generates a drift added to the mid-price after a trade that affects all the future events. It is estimated as the expectation of the long-term impact on the mid-price via a generalized linear model. We specify the function $f$ in equation (\ref{def-f}) as
\[
	f(\delta)=\frac{1}{1+e^{-(B_0 + B_1 \delta)}}
\]
where $B_0$ and $B_1$ are to be estimated {in the logistic regression framework}. The fitted model of INTC is plotted in Figure \ref{fig:logisticFit}, which justifies the general linear relationship between the change of benchmark price and order imbalance. The result also suggests that the probability of a price down-move with more sell orders.  We obtain $\Beta_0=0.0697$, $\Beta_1=0.7624$. The equilibrium order imbalance is then computed as
\[
	\bar{\delta}=\frac{\log(1)-\Beta_0}{\Beta_1}=-0.0915
\]
The average magnitude of a mid-price move is $\bar{Z}=0.5$ ticks; the average size of a market order $\bar{L}=115$ shares. Based on our assumption that mid-prices change at second frequency  the permanent market impact of a sell order is then calculated as %the deviation from the expected martingale price
%\[
%	\bar{\delta}=\frac{\log(1)-\Beta_0}{\Beta_1}=-0.0915
%\]
%\[
%  g(\bar{\delta})=\Beta_0 + \Beta_1*(1+\bar{\delta})=0.7624
%\]
%\[
%  p^{up}=\frac{1}{1+exp(-g(\bar{\delta}))}=0.681
%\]
\[
	\Lambda=0.1821 \hbox{ ticks per average size market order} \quad \mbox{and} \quad  \lambda=\frac{\Lambda}{\bar{L}}=0.0016 \hbox{ ticks per share.}
\]
%\[
%  \lambda=\frac{\Lambda}{\bar{L}}=0.0016 \hbox{ ticks}
%\]
This means on average, we expect one share to cause the expected price change by $0.0016$ ticks. The opening price of INTC for that day is 48.625 dollar a share and the tick size is 0.01, so the permanent market impact factor per share in terms of basis points is 
\[
	\lambda^*=\frac{0.0016*0.01*10000}{48.625}=0.0032 \hbox{ basis points}
\]

%\begin{figure}[!htbp]
%\begin{center}
%\includegraphics[width=\textwidth]{images/logistic_hist.png}
%\end{center}
%\caption{Empirical distribution of order imbalance (buy minus sell) followed by an up move and down move. Large deviations have been truncated for clarity.}
%\label{fig:logisticHist}
%\end{figure}

%%%%%%%%%%%%%%%%%%%%%%%
%%%%%%%%%%%%%%%%%%%%%%%
%%%%%%%%%%%%%%%%%%%%%%%

\subsubsection{Calculation of temporary market impact factor}\label{exmpl_temp}

Temporary market impact gives a temporary drift added to the mid-price after a trade, which decays as time passes. It is estimated as the expected permanent impact of the additional orders arriving under the assumption that market orders follow a Hawkes process with intensity (\ref{Hawkes-intensity}) with exponential kernel and parameters, $\nu$, $A$ and $B$. 

The likelihood function of the simple linear Hawkes process with constant intensity $\nu$ is $L=\prod_i^n \nu e^{-\nu t_n};$ see, for example, \cite{daley2007introductionv1}. For the exponential decay kernel used in this study, the likelihood function is given by $$l=\sum_i^n \log(\nu + A * R(i)) - \nu t_n + \frac{A}{B} \sum^n_{i=1}[e^{-B(t_n - t_i)} - 1]$$ where $R(i) = \sum_{j=1}^{i-1} e^{-B(t_i - t_j)}$; see \cite{Ozaki1979} and \cite{ogata_1981} for details. Since $R(i)$ can be defined recursively as $$R(i) = e^{-B(t_i - t_{i-1})} (1 + R(i-1))$$ the log-likelihood function for the exponential decay Hawkes process is of $O(n)$ instead of $O(n^2)$ complexity as the general Hawkes process. The partial derivatives and the Hessian matrix for the log-likelihood function can also be found in \cite{Ozaki1979}.

The parameters  $\nu$, $A$ and $B$ can be calibrated directly from market data using {the maximum likelihood estimation (MLE)}. The estimated parameters of INTC are reported in Table (\ref{table:HawkesParameters}). Numerical results show that buy and sell order processes have similar intensities, that is, the order flow is symmetric. This is consistent to \cite{almgren2001optimal}. It implies that a trading strategy for execution of a liquidation/sell program is analogous to a buy program. We have $$\frac{A}{B}=\frac{8.647}{10.829}=0.7985.$$ 
Assuming that a trader wants to execute a TWAP strategy for 5.5 hours, from 10:00 a.m to 3:30 p.m Eastern Time the total time in seconds is $T=19800.$ We then have tat 
\[
	\rho=\frac{1-A/B}{T/2} = 0.00002035353 \hbox{/second}
\]
The half-life of the exponential decay of the temporary market impact is 
\[
	t_{\frac{1}{2}}=\frac{\log(2)}{\rho}=\frac{\log(2)}{0.00002035353}=34055.379\hbox{ seconds}=1.720\hbox{ days}
\]
so the temporary market impact caused by us is expected to be half of the initial amount after $1.720$ days.
% The equivalance of $\rho$ in days is:
%\[
%	\rho^{day} = \frac{\log(2)}{t^{day}_{\frac{1}{2}}} = \frac{\log(2)}{1.719} = 0.4
%\]
%or $40\%$/day.
\begin{table}
	\centering
	\begin{tabular}{c|c|c|c}
		\hline
		\hline
		Parameters  & $\nu$ & $\Alpha$ & $\Beta$ \\
		\hline
		Buy orders  & 0.206 & 8.852    & 11.376  \\
		\hline
		Sell orders & 0.196 & 8.647    & 10.829  \\
		\hline
		\hline
	\end{tabular}
	\caption{Parameters for the simple exponential Hawkes processes of market orders based on INTC's data. 
		%The intensity function is $I^m_t = \mu^m + \sum_{t_i < t} \Alpha e^{-\Beta (t-t_i)}$ where $t_i$ are the occurrence times of trades.
	}
	\label{table:HawkesParameters}
\end{table}

%\begin{figure}[!htbp]
%\begin{center}
%\includegraphics[width=\textwidth]{images/gamma_plot.png}
%\end{center}
%\caption{Regression line for instantaneous market impact factor used in Example (\ref{exmpl_inst}). The Pearson's R-square of 82\% suggests that the fit is reasonable, although some tail effects can be observed on the plot.}
%\label{fig:instant_regress}
%\end{figure}

\subsection{Statistical summary of 100 stocks}

We perform the estimation for the 100 NASDAQ stocks to obtain a comprehensive understanding on the possible range of the market impacts. Table (\ref{table:summaryFactors}) lists the statistical summary of the estimated parameters of the 100 stocks, along with the empirical distribution displayed as boxplot in Figure \ref{fig:boxplotImpactFactors}. We observe that:
\begin{itemize}
	\item The instantaneous market impact factor is averaging to about $0.022$ basis points (bps) a share across all stocks, ranging from $0.002$ for Intel Corporation (NASDAQ: INTC) to $0.137$ for M\&T Bank Corporation (NASDAQ: MTB). This implies for each additional share acquires, a market maker demands an additional $0.022$ bps on average.
	\item The permanent market impact factor is averaging to about $0.008$ bps a share across all stocks, ranging from $0.004$ for Microsoft Corporation (NASDAQ: MSFT) to $0.033$ for M\&T Bank Corporation (NASDAQ: MTB). This means for each additional share placed, one expects the mid-price to move by, on average, $0.008$ bps. It also shows the permanent market impact is not as high as the instantaneous market impact. This makes sense since otherwise market makers who absorb the shares would be presented with negative expected profit, due to being adversely selected anytime he or she acquires a share from other participants.
	\item The half-life for the recovery rate is on average $0.608$ days, which in line with empirical research that herding effects tend to be short-term, while splitting effects last for days due to institutional traders unwind their positions \cite{BOUCHAUD200957}. The stock with fastest recovery is Occidental Petroleum Corporation (NASDAQ: OXY) at $0.463$ days, while the slowest is Intel Corporation (NASDAQ: INTC) at $1.720$ days.
\end{itemize}

%%%%%%%%%%%%%%%%%%%%%%%
%%%%%%%%%%%%%%%%%%%%%%%
%%%%%%%%%%%%%%%%%%%%%%%

\section{Order-splitting strategy}\label{secStrategy}

In this section, we consider several order-splitting strategies accounting for different forms of market impacts. We call strategy ALL the optimal strategy derived in Theorem \ref{main1} that comprises all three forms of market impact; we call strategies INS and TMP the benchmark strategies derived in Proposition \ref{main2} when there is only instantaneous impact $(\gamma=0)$ and the benchmark strategy when there is only temporary impact $(\eta \to 0),$ respectively. 

\begin{figure}[!htbp]
	\begin{center}
		\includegraphics[width=\textwidth]{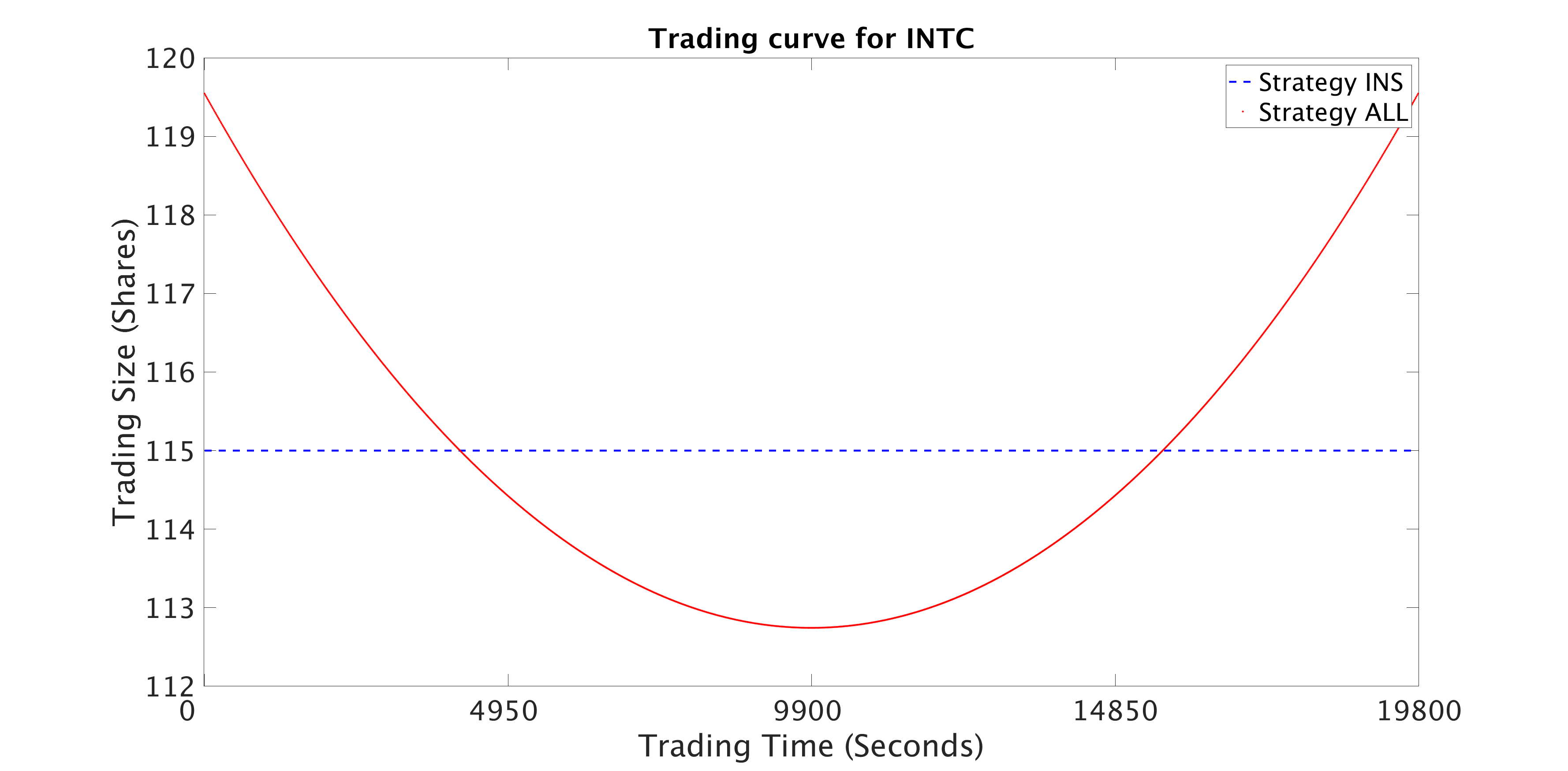}
	\end{center}
	\caption{Trading path of different order-splitting strategies using results in Example (\ref{exmpl_perm}), (\ref{exmpl_temp}) and (\ref{exmpl_inst}). }
	\label{fig:strat_comp}
\end{figure}

Based on the estimated market impact factors, we calibrate the order-splitting strategies using our closed-form formulas. Figure \ref{fig:strat_comp} plots order-splitting strategies ALL and INS according to the parameters calibrated  using the order book of INTC. Strategy ALL increases the trading rate from strategy INS, shifting roughly $14$\% of the total portfolio. As can be seen from the figure, strategy ALL offers a bump of roughly $3.37$\% in trading rate to account for the Hawkes market effects in market order flow. Strategy TMP is comprised of 2 big block trades of size 82,972 shares each, which is approximately $100$ times the liquidity available at the best ask. The theoretical execution costs for strategies INS, ALL and TMP are $3.61$\%, $3.34$\% and $5.12$\% of the portfolio notional, respectively. Strategy ALL seems to be the most reasonable strategy, with expected cost saving of $7.479$\% over strategy INS. 

\begin{figure}[!htbp]
	\begin{center}
		\includegraphics[width=\textwidth]{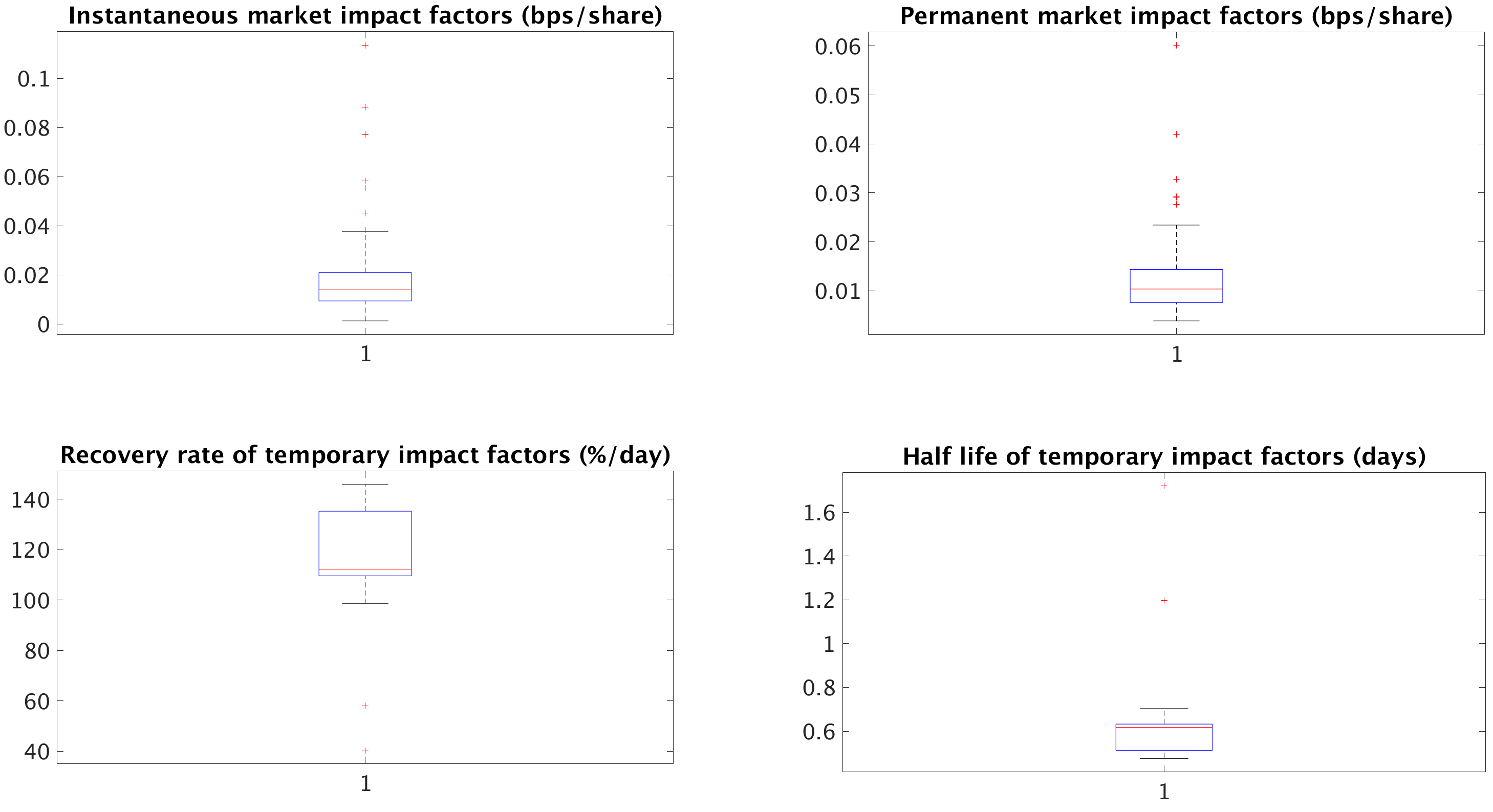}
	\end{center}
	\caption{Summary boxplot for the market impact factors. The recovery rate of the temporary market impact factor is plotted along with its half-life.}
	\label{fig:boxplotImpactFactors}
\end{figure}

We also compare the order splitting strategies for 100 NASDAQ stocks assuming that a trader wants to liquidate a portfolio of $5$\% average daily trading volume. Table (\ref{table:summaryFactors}) list the statistical summary of the estimated parameters of the 100 NASDAQ stocks, along with the empirical distribution displayed as boxplot in Figure \ref{fig:boxplotImpactFactors}. The statistics of the cost differences between the INS and ALL strategies are summarized in Table (\ref{table:summary1}).\footnote{We choose not to perform the comparison for strategy TMP since the block trades calculated are frequently multiplies of current liquidity available in our order book, which will likely to trigger a trading pause or halt. Without certain assumptions on how market react after such events, our results most likely will be inaccurate; see \cite{HAUTSCH2018}, for example, in which the authors investigate how trading pauses affect liquidity.} Strategy ALL outperforms strategy INS, which on average it leads to an improvement of $3.149$\% in terms of trading cost across stocks when using ALL. The lowest one is General Electric Company (NASDAQ: GE) at $1.839$\%, while the best one is Huntington Bancshares Incorporated (NASDAQ: HBAN) at $8.760$\%.

In summary, from both numerical and theoretical standpoints, strategy ALL that takes into account all market impact factors performs better in terms of trading cost across the 100 NASDAQ stocks.

\begin{table}
	\centering
	\begin{tabular}{c|c|c|c|c|c}
		\hline\hline
		                   & Instantaneous & Permanent   & Recovery rate & Half-life & Impact ratio (k) \\
		                   & (bps/share)   & (bps/share) & (\%/day)      & (day)     &                  \\
		\hline
		Min                & 0.002         & 0.001       & 40.295        & 0.463     & 0.129            \\ 
		\hline
		25th percentile    & 0.012         & 0.005       & 108.460       & 0.513     & 0.276            \\ 
		\hline
		Median             & 0.017         & 0.007       & 112.268       & 0.617     & 0.369            \\ 
		\hline
		Mean               & 0.022         & 0.008       & 117.523       & 0.608     & 0.528            \\ 
		\hline
		75th percentile    & 0.027         & 0.010       & 135.242       & 0.639     & 0.552            \\ 
		\hline
		Max                & 0.137         & 0.033       & 149.710       & 1.720     & 4.849            \\
		\hline
		Standard deviation & 0.020         & 0.006       & 17.053        & 0.144     & 0.542            \\
		\hline\hline
	\end{tabular}
	\caption{Summary statistics for the market impact factors. Half-life is the time for the initial expected permanent impact to reduce by half.}
	\label{table:summaryFactors}
\end{table}

\begin{table}
	\centering
	\begin{tabular}{c|c}
		\hline\hline
		                                 & Difference in theoretical execution cost (\%) \\
		\hline
		Min                              & 1.839                                         \\ 
		\hline
		Lower Quartile (25th percentile) & 2.445                                         \\  
		\hline
		Median                           & 2.797                                         \\ 
		\hline
		Mean                             & 3.149                                         \\
		\hline
		Upper Quartile (75th percentile) & 3.417                                         \\  
		\hline
		Max                              & 8.760                                         \\ 
		\hline
		Standard deviation               & 1.068                                         \\ 
		\hline\hline
	\end{tabular}
	\caption{Summary statistics for the cost improvement from using Strategy ALL compared to strategy INS, in percentage of total cost. }
	\label{table:summary1}
\end{table}

%%%%%%%%%%%%%%%%%%%%%%%
%%%%%%%%%%%%%%%%%%%%%%%
%%%%%%%%%%%%%%%%%%%%%%%

\section{Conclusion}\label{secConclusion}

We present three types of market impact that result in different optimal order-splitting strategies, and taking into account all three types of market impact will result in an optimal trading strategy with a closed-form solution. We also detail the procedures of estimating each types of market impact based on market microstructure. Finally, we calculate the theoretical cost savings of the optimal trading strategy that takes into account all types of market impact.

Our model can be extended in various ways to take into account a richer set of empirically observed properties. A richer dynamics of the mid-price process can be modelled by introducing more complex state variables into the generalized linear model. A more complex spread dynamics can also be modelled using a stochastic time process such as GARCH to reflect clustering in time. A more realistic limit order arrival process instead of Poisson birth-death process can also be introduced.

%%%%%%%%%%%%%%%%%%%%%%%
%%%%%%%%%%%%%%%%%%%%%%%
%%%%%%%%%%%%%%%%%%%%%%%

%\clearpage

%\nocite{*}
%\bibliographystyle{siam}
{\small
	\bibliographystyle{agsm}
	\bibliography{references}
}
% Add the References to the table of contents.
%\addcontentsline{toc}{section}{References}

\section{Appendix}\label{secAppendix}

In this appendix we prove Theorem \ref{main1}. The proofs uses variational analysis. The main idea is to characterise the optimal trading strategy as the unique solution to a certain integral equation. For a model with both instantaneous and temporary market impact we show that the integral equation can be reduced to a second order ordinary differential equation (ODE), for which the existence of a classical solution can be established. For a model with only temporary impact, the optimal trading strategy is well known. We also present the derivation of order splitting strategies and numerical results on the 100 NASDAQ stocks.

\subsection{Proof of Theorem \ref{main1}}

In order  to prove our main result, we use a perturbation argument akin to the derivation of the standard Euler-Lagrange equation. For a given deterministic and absolutely continuous portfolio process $X$ with trading rate $\dot X$ the total impact cost is given by
\[
	I(X) =\int_0^T \left(\gamma \dot X_t \int_0^t  \dot X_s e^{-\rho(t-s)}ds  + \eta  \dot X_t^2 \right) dt.
\]
Let ${\cal V}$ be the class of all absolutely continuous deterministic portfolio process $v:[0,T] \to \mathbb R$ with $v(0) = v(T) = 0$ and let $\epsilon >0$.  We expect a stationary point $X^*$ of the the above cost function to satisfy
\[
	\frac{d}{d \epsilon} I(X^* + \epsilon v) |_{\epsilon=0} =0,
\]
for all $v \in {\cal V}$. Standard computations show that this is equivalent to
\[
	\int_0^T v_t \left(\gamma  \int_0^T \dot X^*_s e^{-\rho |t-s| }ds + 2 \eta \dot X^*_t \right) dt = 0,   \quad v \in {\cal V}.
\]
Thus, it follows from the DuBois-Reymend lemma that
\begin{equation}
	\frac{\gamma}{2 \eta}\int_0^T \dot X^*_s e^{-\rho|t-s|} ds + \dot X^*_t = C
\end{equation}
for some constant $C$ and all $t \in [0,T]$. This is a Wiener-Hopf integral equation of the second kind. Differentiating the equation twice w.r.t to the variable $t$ shows that any stationary point of class $C^2$ solves the ODE
\begin{equation}\label{ODE}
	z'' -\rho \left( \frac{\gamma}{\eta} + \rho\right) z = - \rho^2  C
\end{equation}
with boundary conditions
\begin{equation} \label{boundary1}
	\begin{split}
		z_0' =& -\rho( C - \hat \xi_0 )\\
		z_T' =& \rho( C - \hat \xi_T ),
	\end{split}
\end{equation}
where the constant $C$ is determined by the integral condition
\begin{equation} \label{boundary2}
	\int_0^T z_t dt = x.
\end{equation}
The general solution formula for second order ODEs shows that any solution to \eqref{ODE} is of the form
\[
	z_t = C_1 \cosh(kt) +  C_2 \sinh(kt) +  C \cosh(kt) + C\left( 1- \frac{\rho \gamma}{\eta k^2} \right)
\]
for some constants $C_1, C_2$, where $k:= \sqrt{\rho(\rho - \frac{\gamma}{\eta}}$. Inserting conditions \eqref{boundary1} and \eqref{boundary2} yields \eqref{opt}.

It remains to prove that $\xi^*$ is indeed the unique optimal strategy. To this end, we first notice that the mapping $\epsilon \mapsto I(X^* + \epsilon v)$ is defined on the whole real line, is twice continuously differentiable, and has a critical point in $\epsilon = 0$. The second derivative
\[
	\frac{d^2}{d \epsilon^2} I(X^* + \epsilon v) = 2 \gamma \int_0^T \dot v_t \int_0^t \dot v_s e^{-\rho(t-s)} ds \, dt + 2 \eta \int_0^T \dot v^2_t dt 
\]
is strictly positive because the right hand side of the above equation equals twice the cost associated with the trading strategy $\dot v$. In particular, $\epsilon \mapsto I(X^* + \epsilon v)$ is strictly convex for any $v \in {\cal V}$. We conclude that $\xi^*$ is indeed the unique optimal trading strategy.

%%%%%%%%%%%%%%%%%%%%%%%%%%
%%%%%%%%%%%%%%%%%%%%%%%%%%
%%%%%%%%%%%%%%%%%%%%%%%%%%

\subsection{Numerical results}
In this section, we report the detailed numerical results on the 100 NASDAQ stocks.
	{\small \begin{table}\label{table1}
		\centering
		\begin{tabular}{r|r|r|r|r|r|r}
			\hline
			\hline
			Index & Stock Code & Sector           & Market cap & Number of events & Ratio & Cost difference (\%) \\ \hline
			1     & FDX        & Transportation   & 62.474     & 78734            & 3     & 3.410                \\ 
			\hline 
			2     & MSFT       & Technology       & 816.493    & 1549182          & 54    & 3.742                \\
			\hline 
			3     & INTC       & Technology       & 242.087    & 1141474          & 40    & 7.479                \\
			\hline 
			4     & ORCL       & Technology       & 193.166    & 505411           & 18    & 3.492                \\ 
			\hline 
			5     & IBM        & Technology       & 134.345    & 230050           & 8     & 2.238                \\ 
			\hline 
			6     & TXN        & Technology       & 112.546    & 374627           & 13    & 2.275                \\
			\hline 
			7     & QCOM       & Technology       & 86.910     & 242692           & 8     & 3.109                \\ 
			\hline 
			8     & INTU       & Technology       & 55.238     & 84519            & 3     & 2.451                \\
			\hline 
			9     & ITW        & Technology       & 49.751     & 50518            & 2     & 3.261                \\ 
			\hline 
			10    & HPQ        & Technology       & 36.945     & 300305           & 10    & 4.620                \\
			\hline 
			11    & LRCX       & Technology       & 29.177     & 176146           & 6     & 2.305                \\
			\hline 
			12    & HPE        & Technology       & 23.421     & 151653           & 5     & 4.929                \\ 
			\hline 
			13    & INFO       & Technology       & 20.667     & 194728           & 7     & 3.325                \\
			\hline 
			14    & LLL        & Technology       & 15.975     & 33836            & 1     & 2.192                \\
			\hline 
			15    & IPG        & Technology       & 8.281      & 141214           & 5     & 3.155                \\
			\hline 
			16    & T          & Public Utilities & 227.848    & 394611           & 14    & 4.285                \\ 
			\hline 
			17    & VZ         & Public Utilities & 209.157    & 398198           & 14    & 4.949                \\
			\hline 
			18    & NEE        & Public Utilities & 80.130     & 40624            & 1     & 3.560                \\
			\hline 
			19    & SO         & Public Utilities & 48.072     & 254492           & 9     & 2.456                \\
			\hline 
			20    & EXC        & Public Utilities & 40.594     & 217442           & 8     & 3.315                \\ 
			\hline 
			21    & KMI        & Public Utilities & 39.025     & 247063           & 9     & 2.601                \\ 
			\hline 
			22    & V          & Miscellaneous    & 325.321    & 285520           & 10    & 2.347                \\
			\hline 
			23    & ACN        & Miscellaneous    & 106.808    & 55501            & 2     & 2.477                \\
			\hline 
			24    & PYPL       & Miscellaneous    & 103.884    & 360626           & 12    & 2.617                \\ 
			\hline 
			25    & JNJ        & Health Care      & 337.549    & 268100           & 9     & 2.609                \\ 
			\hline 
			26    & PFE        & Health Care      & 218.364    & 474758           & 16    & 2.109                \\
			\hline 
			27    & MRK        & Health Care      & 168.198    & 399463           & 14    & 3.424                \\
			\hline 
			28    & MMM        & Health Care      & 119.896    & 117872           & 4     & 3.826                \\ 
			\hline 
			29    & MDT        & Health Care      & 118.977    & 206328           & 7     & 2.992                \\ 
			\hline 
			30    & BMY        & Health Care      & 92.695     & 189814           & 7     & 2.323                \\
			\hline 
			31    & BIIB       & Health Care      & 75.646     & 73727            & 3     & 2.915                \\
			\hline 
			32    & WBA        & Health Care      & 64.482     & 351245           & 12    & 2.036                \\ 
			\hline 
			33    & ISRG       & Health Care      & 58.568     & 55673            & 2     & 2.320                \\
			\hline 
			34    & AGN        & Health Care      & 58.248     & 53729            & 2     & 2.211                \\ 
			\hline 
			35    & INCY       & Health Care      & 14.670     & 35050            & 1     & 2.489                \\
			\hline 
			36    & HOLX       & Health Care      & 11.223     & 168340           & 6     & 2.727                \\ 
			\hline 
			37    & JPM        & Finance          & 374.001    & 778145           & 27    & 3.301                \\
			\hline 
			38    & BAC        & Finance          & 301.683    & 829884           & 29    & 2.808                \\
			\hline 
			39    & C          & Finance          & 174.199    & 823890           & 29    & 4.595                \\
			\hline 
			40    & MS         & Finance          & 87.972     & 452470           & 16    & 3.097                \\
			\hline 
			41    & GS         & Finance          & 87.514     & 94318            & 3     & 4.507                \\
			\hline 
			42    & AXP        & Finance          & 86.165     & 140248           & 5     & 2.863                \\
			\hline 
			43    & USB        & Finance          & 83.864     & 371890           & 13    & 3.339                \\ 
			\hline 
			44    & BK         & Finance          & 53.020     & 262987           & 9     & 2.713                \\ 
			\hline 
			45    & AIG        & Finance          & 47.864     & 253522           & 9     & 2.786                \\
			\hline 
			46    & COF        & Finance          & 47.296     & 102156           & 4     & 2.477                \\
			\hline 
			47    & MET        & Finance          & 44.483     & 510715           & 18    & 3.143                \\ 
			\hline 
			48    & ICE        & Finance          & 43.911     & 118776           & 4     & 2.302                \\
			\hline 
			49    & ALL        & Finance          & 32.836     & 101674           & 4     & 2.699                \\
			\hline 
			50    & MTB        & Finance          & 25.126     & 34463            & 1     & 2.293                \\ 
			\hline
			\hline
		\end{tabular}
		{{\caption{\footnotesize{Reference table of stock codes, sectors, market cap in billion(s) of USD, total event count and cost improvements (from index 1 to 50). The Ratio column reports the ratio in percentage of number of total events compared to SJM. The last column reports cost improvements by using Strategy ALL compared to Strategy INS.}}}}
		\label{table:StockCodeTable1}
	\end{table}

\begin{table} \label{table2}
	\centering
	\begin{tabular}{r|r|r|r|r|r|r}
		\hline
		\hline
		Index & Stock Code & Sector                & Market cap & Number of events & Ratio & Cost difference (\%) \\ \hline
		51    & KEY        & Finance               & 21.785     & 325154           & 11    & 4.588                \\ 
		\hline 
		52    & HBAN       & Finance               & 16.637     & 312042           & 11    & 8.760                \\
		\hline 
		53    & L          & Finance               & 15.845     & 114019           & 4     & 2.499                \\
		\hline 
		54    & IVZ        & Finance               & 10.458     & 296285           & 10    & 2.938                \\ 
		\hline 
		55    & GE         & Energy                & 113.952    & 407933           & 14    & 1.839                \\
		\hline 
		56    & SLB        & Energy                & 91.543     & 260220           & 9     & 3.409                \\ 
		\hline 
		57    & COP        & Energy                & 81.916     & 260091           & 9     & 2.190                \\ 
		\hline 
		58    & OXY        & Energy                & 63.620     & 250987           & 9     & 2.238                \\ 
		\hline 
		59    & EMR        & Energy                & 43.653     & 119474           & 4     & 3.132                \\ 
		\hline 
		60    & HAL        & Energy                & 39.594     & 318400           & 11    & 4.663                \\
		\hline 
		61    & MPC        & Energy                & 33.077     & 147308           & 5     & 3.167                \\ 
		\hline 
		62    & HES        & Energy                & 19.219     & 138985           & 5     & 2.668                \\ 
		\hline 
		63    & MRO        & Energy                & 17.473     & 285446           & 10    & 4.441                \\
		\hline 
		64    & AMZN       & Consumer Services     & 880.260    & 196082           & 7     & 3.703                \\
		\hline 
		65    & HD         & Consumer Services     & 233.535    & 264433           & 9     & 2.623                \\
		\hline 
		66    & CMCSA      & Consumer Services     & 157.882    & 644157           & 22    & 6.495                \\
		\hline 
		67    & MCD        & Consumer Services     & 124.034    & 229323           & 8     & 2.844                \\ 
		\hline 
		68    & COST       & Consumer Services     & 95.918     & 129075           & 4     & 2.251                \\ 
		\hline 
		69    & LOW        & Consumer Services     & 82.154     & 258685           & 9     & 2.257                \\
		\hline 
		70    & SBUX       & Consumer Services     & 70.256     & 417836           & 14    & 4.653                \\
		\hline 
		71    & TGT        & Consumer Services     & 41.457     & 185869           & 6     & 2.462                \\
		\hline 
		72    & HLT        & Consumer Services     & 24.622     & 67404            & 2     & 2.741                \\ 
		\hline 
		73    & KSS        & Consumer Services     & 12.307     & 125230           & 4     & 2.500                \\
		\hline 
		74    & HCP        & Consumer Services     & 12.205     & 139467           & 5     & 4.208                \\
		\hline 
		75    & M          & Consumer Services     & 11.841     & 342265           & 12    & 3.872                \\ 
		\hline 
		76    & LB         & Consumer Services     & 8.787      & 166285           & 6     & 3.191                \\
		\hline 
		77    & HBI        & Consumer Services     & 7.989      & 139628           & 5     & 4.286                \\ 
		\hline 
		78    & KIM        & Consumer Services     & 7.075      & 112259           & 4     & 3.038                \\ 
		\hline 
		79    & KO         & Consumer Non-Durables & 192.678    & 302881           & 10    & 5.314                \\ 
		\hline 
		80    & PEP        & Consumer Non-Durables & 164.075    & 250050           & 9     & 2.383                \\ 
		\hline 
		81    & NKE        & Consumer Non-Durables & 124.041    & 393240           & 14    & 3.206                \\
		\hline 
		82    & MO         & Consumer Non-Durables & 109.092    & 282687           & 10    & 2.809                \\ 
		\hline 
		83    & K          & Consumer Non-Durables & 24.349     & 107317           & 4     & 2.348                \\
		\hline 
		84    & SJM        & Consumer Non-Durables & 12.429     & 28872            & 1     & 2.351                \\ 
		\hline 
		85    & HAS        & Consumer Non-Durables & 12.025     & 46148            & 2     & 2.975                \\ 
		\hline 
		86    & HOG        & Consumer Non-Durables & 6.927      & 114499           & 4     & 2.520                \\ 
		\hline 
		87    & LEG        & Consumer Durables     & 6.045      & 46027            & 2     & 2.398                \\
		\hline 
		88    & BA         & Capital Goods         & 206.758    & 105923           & 4     & 3.274                \\
		\hline 
		89    & HON        & Capital Goods         & 114.377    & 140798           & 5     & 2.693                \\ 
		\hline 
		90    & LMT        & Capital Goods         & 92.098     & 63061            & 2     & 2.381                \\ 
		\hline 
		91    & CAT        & Capital Goods         & 81.811     & 261324           & 9     & 3.656                \\
		\hline 
		92    & GD         & Capital Goods         & 58.005     & 54130            & 2     & 2.615                \\
		\hline 
		93    & RTN        & Capital Goods         & 57.864     & 35425            & 1     & 2.528                \\ 
		\hline 
		94    & ILMN       & Capital Goods         & 45.746     & 39506            & 1     & 2.770                \\
		\hline 
		95    & F          & Capital Goods         & 42.081     & 230713           & 8     & 5.219                \\
		\hline 
		96    & PG         & Basic Industries      & 197.849    & 374423           & 13    & 2.434                \\
		\hline 
		97    & LYB        & Basic Industries      & 42.489     & 99786            & 3     & 2.244                \\
		\hline 
		98    & IP         & Basic Industries      & 21.881     & 73085            & 3     & 2.536                \\ 
		\hline 
		99    & LEN        & Basic Industries      & 18.016     & 238777           & 8     & 2.548                \\ 
		\hline 
		100   & JEC        & Basic Industries      & 9.371      & 66381            & 2     & 2.440                \\ 
		\hline
		\hline
	\end{tabular}
	{{\caption{\footnotesize{Reference table of stock codes, sectors, market cap in billion(s) of USD, total event count and cost improvements (from index 51 to 100). The Ratio column reports the ratio in percentage of number of total events compared to SJM. The last column reports cost improvements by using Strategy ALL compared to Strategy INS.}}}}
	\label{table:StockCodeTable2}
\end{table}}

\begin{table}
	\centering
	\begin{tabular}{r|r|r|r|r|r|r}
		\hline
		\hline
		   & Stock Code & Instantaneous & Permanent   & Recovery rate & Half-life & Impact ratio \\
		   &            & (bps/share)   & (bps/share) & (\%/day)      & (day)     &              \\
		\hline
		1  & FDX        & 0.016         & 0.009       & 109.619       & 0.632     & 0.549        \\ 
		\hline 
		2  & MSFT       & 0.004         & 0.003       & 114.715       & 0.604     & 0.658        \\
		\hline 
		3  & INTC       & 0.002         & 0.003       & 40.295        & 1.720     & 1.450        \\ 
		\hline 
		4  & ORCL       & 0.008         & 0.004       & 98.576        & 0.703     & 0.576        \\ 
		\hline 
		5  & IBM        & 0.020         & 0.004       & 135.242       & 0.513     & 0.224        \\ 
		\hline 
		6  & TXN        & 0.025         & 0.006       & 98.576        & 0.703     & 0.233        \\
		\hline 
		7  & QCOM       & 0.012         & 0.005       & 129.922       & 0.534     & 0.458        \\
		\hline 
		8  & INTU       & 0.019         & 0.005       & 109.619       & 0.632     & 0.277        \\
		\hline 
		9  & ITW        & 0.015         & 0.007       & 111.191       & 0.623     & 0.503        \\ 
		\hline 
		10 & HPQ        & 0.009         & 0.009       & 111.191       & 0.623     & 0.991        \\
		\hline 
		11 & LRCX       & 0.035         & 0.008       & 107.302       & 0.646     & 0.240        \\ 
		\hline 
		12 & HPE        & 0.017         & 0.019       & 113.888       & 0.609     & 1.128        \\
		\hline 
		13 & INFO       & 0.014         & 0.007       & 138.663       & 0.500     & 0.523        \\ 
		\hline 
		14 & LLL        & 0.046         & 0.010       & 101.443       & 0.683     & 0.212        \\
		\hline 
		15 & IPG        & 0.018         & 0.009       & 112.268       & 0.617     & 0.471        \\
		\hline 
		16 & T          & 0.005         & 0.004       & 109.619       & 0.632     & 0.856        \\
		\hline 
		17 & VZ         & 0.004         & 0.004       & 103.845       & 0.667     & 1.137        \\ 
		\hline 
		18 & NEE        & 0.024         & 0.014       & 111.191       & 0.623     & 0.597        \\
		\hline 
		19 & SO         & 0.019         & 0.005       & 109.619       & 0.632     & 0.278        \\ 
		\hline 
		20 & EXC        & 0.015         & 0.008       & 109.619       & 0.632     & 0.520        \\
		\hline 
		21 & KMI        & 0.023         & 0.007       & 113.012       & 0.613     & 0.316        \\
		\hline 
		22 & V          & 0.017         & 0.004       & 129.922       & 0.534     & 0.251        \\
		\hline 
		23 & ACN        & 0.028         & 0.008       & 109.619       & 0.632     & 0.284        \\
		\hline 
		24 & PYPL       & 0.015         & 0.005       & 121.293       & 0.571     & 0.320        \\ 
		\hline 
		25 & JNJ        & 0.013         & 0.004       & 144.621       & 0.479     & 0.318        \\
		\hline 
		26 & PFE        & 0.014         & 0.003       & 107.302       & 0.646     & 0.192        \\
		\hline 
		27 & MRK        & 0.008         & 0.004       & 135.242       & 0.513     & 0.554        \\
		\hline 
		28 & MMM        & 0.007         & 0.005       & 138.663       & 0.500     & 0.687        \\
		\hline 
		29 & MDT        & 0.015         & 0.006       & 135.242       & 0.513     & 0.424        \\
		\hline 
		30 & BMY        & 0.027         & 0.007       & 135.242       & 0.513     & 0.245        \\
		\hline 
		31 & BIIB       & 0.040         & 0.016       & 111.191       & 0.623     & 0.402        \\ 
		\hline 
		32 & WBA        & 0.025         & 0.004       & 121.293       & 0.571     & 0.175        \\
		\hline 
		33 & ISRG       & 0.098         & 0.024       & 112.268       & 0.617     & 0.244        \\ 
		\hline 
		34 & AGN        & 0.083         & 0.018       & 111.191       & 0.623     & 0.217        \\
		\hline 
		35 & INCY       & 0.045         & 0.013       & 112.268       & 0.617     & 0.287        \\ 
		\hline 
		36 & HOLX       & 0.017         & 0.006       & 109.619       & 0.632     & 0.350        \\ 
		\hline 
		37 & JPM        & 0.006         & 0.003       & 145.827       & 0.475     & 0.515        \\
		\hline 
		38 & BAC        & 0.004         & 0.002       & 103.845       & 0.667     & 0.372        \\
		\hline 
		39 & C          & 0.004         & 0.004       & 136.246       & 0.509     & 0.981        \\
		\hline 
		40 & MS         & 0.016         & 0.007       & 144.621       & 0.479     & 0.454        \\ 
		\hline 
		41 & GS         & 0.015         & 0.014       & 103.845       & 0.667     & 0.944        \\ 
		\hline 
		42 & AXP        & 0.017         & 0.007       & 138.663       & 0.500     & 0.387        \\
		\hline 
		43 & USB        & 0.010         & 0.005       & 135.242       & 0.513     & 0.527        \\
		\hline 
		44 & BK         & 0.019         & 0.006       & 135.242       & 0.513     & 0.346        \\ 
		\hline 
		45 & AIG        & 0.016         & 0.006       & 135.242       & 0.513     & 0.366        \\
		\hline 
		46 & COF        & 0.046         & 0.013       & 107.302       & 0.646     & 0.284        \\
		\hline 
		47 & MET        & 0.010         & 0.005       & 129.922       & 0.534     & 0.468        \\
		\hline 
		48 & ICE        & 0.035         & 0.008       & 138.663       & 0.500     & 0.239        \\
		\hline 
		49 & ALL        & 0.027         & 0.009       & 109.619       & 0.632     & 0.342        \\
		\hline 
		50 & MTB        & 0.137         & 0.033       & 113.529       & 0.611     & 0.237        \\    
		\hline
		\hline
	\end{tabular}
	\caption{Market impact factors estimation, from index 1 to 50.}
	\label{table:impactFactor1}
\end{table}
\begin{table}
	\centering
	\begin{tabular}{r|r|r|r|r|r|r}
		\hline
		\hline
		    & Stock Code & Instantaneous & Permanent   & Recovery rate & Half-life & Impact ratio \\
		    &            & (bps/share)   & (bps/share) & (\%/day)      & (day)     &              \\    
		\hline
		51  & KEY        & 0.016         & 0.016       & 111.191       & 0.623     & 0.978        \\ 
		\hline
		52  & HBAN       & 0.003         & 0.014       & 113.012       & 0.613     & 4.849        \\
		\hline
		53  & L          & 0.023         & 0.007       & 138.663       & 0.500     & 0.289        \\ 
		\hline
		54  & IVZ        & 0.027         & 0.011       & 107.302       & 0.646     & 0.408        \\ 
		\hline
		55  & GE         & 0.011         & 0.001       & 113.529       & 0.611     & 0.129        \\
		\hline
		56  & SLB        & 0.012         & 0.007       & 129.922       & 0.534     & 0.549        \\ 
		\hline
		57  & COP        & 0.028         & 0.006       & 129.922       & 0.534     & 0.212        \\ 
		\hline
		58  & OXY        & 0.027         & 0.006       & 149.710       & 0.463     & 0.224        \\ 
		\hline
		59  & EMR        & 0.012         & 0.006       & 138.663       & 0.500     & 0.465        \\ 
		\hline
		60  & HAL        & 0.006         & 0.007       & 129.922       & 0.534     & 1.010        \\ 
		\hline
		61  & MPC        & 0.015         & 0.007       & 103.845       & 0.667     & 0.475        \\ 
		\hline
		62  & HES        & 0.036         & 0.012       & 138.663       & 0.500     & 0.334        \\ 
		\hline
		63  & MRO        & 0.015         & 0.014       & 109.619       & 0.632     & 0.917        \\ 
		\hline
		64  & AMZN       & 0.015         & 0.010       & 57.959        & 1.196     & 0.645        \\ 
		\hline
		65  & HD         & 0.014         & 0.005       & 103.845       & 0.667     & 0.322        \\
		\hline
		66  & CMCSA      & 0.005         & 0.010       & 103.845       & 0.667     & 2.046        \\ 
		\hline
		67  & MCD        & 0.009         & 0.003       & 103.845       & 0.667     & 0.382        \\ 
		\hline
		68  & COST       & 0.025         & 0.006       & 107.302       & 0.646     & 0.227        \\
		\hline
		69  & LOW        & 0.022         & 0.005       & 144.621       & 0.479     & 0.228        \\ 
		\hline
		70  & SBUX       & 0.004         & 0.004       & 98.576        & 0.703     & 1.005        \\ 
		\hline
		71  & TGT        & 0.015         & 0.004       & 135.242       & 0.513     & 0.280        \\ 
		\hline
		72  & HLT        & 0.021         & 0.007       & 109.619       & 0.632     & 0.353        \\ 
		\hline
		73  & KSS        & 0.029         & 0.008       & 138.663       & 0.500     & 0.290        \\ 
		\hline
		74  & HCP        & 0.022         & 0.018       & 112.268       & 0.617     & 0.826        \\ 
		\hline
		75  & M          & 0.013         & 0.009       & 107.302       & 0.646     & 0.703        \\ 
		\hline
		76  & LB         & 0.023         & 0.011       & 138.663       & 0.500     & 0.482        \\
		\hline
		77  & HBI        & 0.034         & 0.029       & 111.191       & 0.623     & 0.856        \\ 
		\hline
		78  & KIM        & 0.024         & 0.011       & 113.012       & 0.613     & 0.437        \\ 
		\hline
		79  & KO         & 0.003         & 0.004       & 109.619       & 0.632     & 1.315        \\
		\hline
		80  & PEP        & 0.014         & 0.004       & 98.576        & 0.703     & 0.260        \\
		\hline
		81  & NKE        & 0.009         & 0.005       & 129.922       & 0.534     & 0.487        \\ 
		\hline
		82  & MO         & 0.014         & 0.005       & 129.922       & 0.534     & 0.372        \\
		\hline
		83  & K          & 0.024         & 0.006       & 107.302       & 0.646     & 0.251        \\ 
		\hline
		84  & SJM        & 0.087         & 0.022       & 112.268       & 0.617     & 0.252        \\ 
		\hline
		85  & HAS        & 0.017         & 0.007       & 112.268       & 0.617     & 0.419        \\ 
		\hline
		86  & HOG        & 0.024         & 0.007       & 140.927       & 0.492     & 0.295        \\ 
		\hline
		87  & LEG        & 0.045         & 0.012       & 111.191       & 0.623     & 0.264        \\ 
		\hline
		88  & BA         & 0.016         & 0.008       & 103.845       & 0.667     & 0.507        \\
		\hline
		89  & HON        & 0.019         & 0.006       & 107.302       & 0.646     & 0.341        \\ 
		\hline
		90  & LMT        & 0.030         & 0.008       & 111.191       & 0.623     & 0.259        \\ 
		\hline
		91  & CAT        & 0.010         & 0.006       & 135.242       & 0.513     & 0.629        \\ 
		\hline
		92  & GD         & 0.021         & 0.007       & 107.302       & 0.646     & 0.320        \\
		\hline
		93  & RTN        & 0.027         & 0.008       & 112.268       & 0.617     & 0.297        \\ 
		\hline
		94  & ILMN       & 0.061         & 0.022       & 112.268       & 0.617     & 0.362        \\
		\hline
		95  & F          & 0.002         & 0.003       & 114.517       & 0.605     & 1.267        \\ 
		\hline
		96  & PG         & 0.010         & 0.003       & 129.922       & 0.534     & 0.273        \\
		\hline
		97  & LYB        & 0.044         & 0.010       & 107.302       & 0.646     & 0.225        \\ 
		\hline
		98  & IP         & 0.028         & 0.008       & 111.191       & 0.623     & 0.299        \\
		\hline
		99  & LEN        & 0.017         & 0.005       & 136.246       & 0.509     & 0.302        \\ 
		\hline
		100 & JEC        & 0.037         & 0.010       & 109.619       & 0.632     & 0.274        \\
		\hline
		\hline
	\end{tabular}
	\caption{Market impact factors estimation, from index 51 to 100}
	\label{table:impactFactor2}
\end{table}

\end{document}